# Self-Dual Supersymmetric Dirac-Born-Infeld Action in Three-Dimensions

Hitoshi Nishino,[1] Subhash Rajpoot[2] and Kevin Reed

*Department of Physics & Astronomy*
*California State University*
*1250 Bellflower Boulevard*
*Long Beach, CA 90840*

## Abstract

We present a self-dual $N=1$ supersymmetric Dirac-Born-Infeld action in three dimensions. This action is based on the supersymmetric generalized self-duality in odd dimensions developed originally by Townsend, Pilch and van Nieuwenhuizen. Even though such a self-duality had been supposed to be very difficult to generalize to a supersymmetrically interacting system, we show that Dirac-Born-Infeld action is actually compatible with supersymmetry and self-duality in three-dimensions. The interactions can be further generalized to arbitrary (non)polynomial interactions. As a by-product, we also show that a third-rank field strength leads to a more natural formulation of self-duality in 3D. We also show an interesting role played by the third-rank field strength leading to a supersymmetry breaking, in addition to accommodating a Chern-Simons form.

PACS: 12.60.Jv, 11.30.Pb, 11.10.Lm, 11.25.Tq

Key Words: Supersymmetry, Dirac-Born-Infeld, Non-Linear Electrodynamics, Self-Duality, Three-Dimensions

---

[1]E-Mail: hnishino@csulb.edu
[2]E-Mail: rajpoot@csulb.edu

1. **Introduction**

The original nonlinear electrodynamics by Born and Infeld in 1930's [1] is drawing more attention nowadays, after the importance of non-linear electrodynamics has been recognized in the context of superstring theories [2]. Moreover, it has been widely realized that soliton-like solutions are very important in Dp-branes described in terms of Dirac-Born-Infeld (DBI) like actions [3]. These nonlinear theories have been been also generalized by supersymmetry [4] or non-Abelian gauge groups [5].

Independent of these, there has been a development related to 'self-duality' in odd-dimensions, as a generalization of Hodge self-duality in even dimensions. For example, self-duality in three-dimensions (3D) was first considered in [6], while in 7D it was developed as a solution to a problem with gauge covariant field strengths for antisymmetric tensors [7]. This solution came from the $S^4$-compactification of 11D supergravity into 7D, leading to a massive self-dual 3-form tensor field in 7D [7]. The generalized self-duality in 3D [7] can be defined as follows: Consider an Abelian vector field $A_a$ with the lagrangian[3]

$$\mathcal{L}_{\text{SD}} \equiv -\tfrac{1}{2} m^2 A_a A^a + \tfrac{1}{4} m \epsilon^{abc} A_a F_{bc} \ , \tag{1.1}$$

where the first term is the usual mass term for the vector field, while the second term is an Abelian Chern-Simons (CS) term. Interestingly, the field equation for $A_a$ is

$$A_a \doteq +\tfrac{1}{2} m^{-1} \epsilon_a{}^{bc} F_{bc} \ , \tag{1.2}$$

which seems to be a generalization of 'self-duality' in even dimensions, with the usual field strength on the l.h.s. replaced by the potential $A_a$.

As has been shown in [7], this system has only one propagating degree of freedom out of $A_a$. In other words, the system (1.2) is a 'square root' of the usual massive case $\partial_b F_a{}^b \doteq m^2 A_a$ with two propagating degrees of freedom. Note also that this propagation is realized *without* a kinetic term of the $F_{ab}{}^2$-type in (1.1). We can also supersymmetrize the lagrangian (1.1), by supplying a massive Majorana fermion $\lambda_\alpha$ with one degree of freedom [7]. There have been also recent developments about the duality equivalence between 'topological massive theory' with the terms $F_{ab}{}^2 \oplus m\epsilon^{abc} A_a F_{bc}$ and self-dual theory above [9], or the compatibility between self-duality and non-linear electrodynamics has been studied for purely bosonic system [10]. Additionally, the wide equivalence between these gauge models has been recently pointed out [11].

In this paper, we will combine the three concepts related to vector multiplets in 3D, *i.e.,* the non-linear electrodynamics such as DBI action, the generalized self-duality in 3D, and supersymmetry. We show that these three concepts are indeed compatible with each other, and in particular, we will confirm it in terms of superspace language. Our results have some overlap with some recent results related to supersymmetric DBI actions, such as that in [12]. However, we stress that the so-called self-duality in 3D requires the non-trivial involvement of component fields that are usually absent in the Wess-Zumino gauge. We show that despite the involvement of these fields, the $N=1$ supersymmetrization of self-duality in 3D is compatible with nontrivial interactions such as DBI actions. In other words, we will accomplish the supersymmetrization of self-dual non-linear electrodynamics similar to [10]. As a by-product, we will also show that introducing a third-rank field strength is more natural for formulating self-duality in 3D.

---

[3]We use the notation in [8], namely, we use the signature $(\eta_{ab}) = \text{diag.}(-,+,+)$, where $a, b, \cdots = 0, 1, 2$ for vector indices in 3D, while $\alpha, \beta, \cdots = 1, 2$ are used for Majorana spinors in 3D.

## 2. Total Action

We have a simple $N=1$ vector multiplet field content $(\chi_\alpha, B, V_{\alpha\beta}, \lambda_\alpha)$, where we adopt the superspace notation in 3D in [8], namely, $\chi_\alpha$ ($\alpha, \beta, \cdots = 1, 2$) is a Majorana spinor, $B$ is a real scalar, $V_{\alpha\beta}$ is a real vector field in terms of symmetric spinorial indices, while $\lambda_\alpha$ is a two-component Majorana spinor field. In the usual Wess-Zumino gauge, the fields $\chi_\alpha$ and $B$ can be completely gauged away, because no terms involving them are needed. However, in our self-duality formulation, we need them due to the broken Abelian gauge symmetry, reflected by the presence of the mass term in (1.2). The component fields above are related to the spinor superfield $\Gamma_\alpha$ as [8]

$$\Gamma_\alpha| = \chi_\alpha \ , \quad \tfrac{1}{2}D^\alpha \Gamma_\alpha| = B \ , \tag{2.1a}$$

$$W_\alpha \equiv \tfrac{1}{2}D^\beta D_\alpha \Gamma_\beta \ , \quad W_\alpha| = \lambda_\alpha \tag{2.1b}$$

$$\Gamma_{\alpha\beta} \equiv -\tfrac{i}{2} D_{(\alpha}\Gamma_{\beta)} \ , \quad \Gamma_{\alpha\beta}| = V_{\alpha\beta} \ , \tag{2.1c}$$

$$D_\alpha W_\beta = D_\beta W_\alpha \ , \quad D_\alpha W_\beta| = f_{\alpha\beta} \ , \tag{2.1d}$$

based on the most fundamental superspace relationship [8]

$$\{D_\alpha, D_\beta\} = +2i\partial_{\alpha\beta} \ . \tag{2.2}$$

Our total action $I_{\text{tot}}$ is conveniently given in terms of three actions $I_T$, $I_{m^2}$ and $I_{\text{CS}}$:[4]

$$I_{\text{tot}} \equiv I_T + I_{m^2} + I_{\text{CS}} \ , \tag{2.3a}$$

$$I_T \equiv \int d^3x\, d^2\theta \left[ T(D^2 W^2) \cdot W^2 + \tfrac{c^2}{2} W^2 \right] \equiv \int d^3x\, d^2\theta\, \mathcal{L}_T(z) \equiv \int d^3x\, \mathcal{L}_T(x)$$

---
[4]We use the dot-symbol $\cdot$ in these expressions to stress the difference of a product from the variable $D^2 W^2$ of the function $T(D^2 W^2)$.

$$= \int d^3x \left[ T(D^2W^2) \cdot D^2W^2 + T'(D^2W^2) \cdot \left(\tfrac{1}{2}W^2 \partial_{\alpha\beta}\partial^{\alpha\beta}W^2 + iP^\alpha \partial_\alpha{}^\beta P_\beta\right) \right.$$
$$\left. - \tfrac{1}{2}T''(D^2W^2) \cdot (\partial^{\alpha\beta}P_\beta)(\partial_\alpha{}^\gamma P_\gamma) \cdot W^2 + \tfrac{c^2}{2}D^2(W^2) \right] \Big| \quad , \tag{2.3b}$$

$$I_{m^2} \equiv \int d^3x\, d^2\theta\, \tfrac{1}{2}am^2\,(\Gamma^\alpha \Gamma_\alpha) \equiv \int d^3x\, d^2\theta\, \mathcal{L}_{m^2}(z) \equiv \int d^3x\, \mathcal{L}_{m^2}(x)$$
$$= \int d^3x\, a\left(+2m^2 \lambda^\alpha \chi_\alpha - im^2 \chi_\alpha \partial_\alpha{}^\beta \chi_\beta + \tfrac{1}{2}m^2 V_{\alpha\beta} V^{\alpha\beta} - m^2 B^2\right) \quad , \tag{2.3c}$$

$$I_{\rm CS} \equiv \int d^3x\, d^2\theta\, \tfrac{1}{2}bm\, \Gamma^\alpha W_\alpha \equiv \int d^3x\, d^2\theta\, \mathcal{L}_{\rm CS}(z) \equiv \int d^3x\, \mathcal{L}_{\rm CS}(x)$$
$$= \int d^3x\, b\left(-\tfrac{i}{4}mV^{\alpha\beta}f_{\alpha\beta} + m\lambda^\alpha \lambda_\alpha\right) \quad . \tag{2.3d}$$

The $a$ and $b$ are non-zero dimensionless real constants for normalizations, while the constant $m$ has the dimension of mass. The $D^2 \equiv (1/2)D^\alpha D_\alpha$ complies with the notation in [8]. Our useful relations and symbols are

$$\mathcal{L}_0(x) \equiv D^2(W^2)\big| \equiv -\tfrac{1}{2}f_{\alpha\beta}f^{\alpha\beta} + i\lambda^\alpha \partial_\alpha{}^\beta \lambda_\beta \quad , \tag{2.4a}$$

$$P_\alpha \equiv D_a(W^2) = f_\alpha{}^\beta W_\beta \quad . \tag{2.4b}$$

The real function $T(D^2W^2)$ is *a priori* a general finite polynomial or infinite power series in terms of $D^2W^2$. The *prime* symbol, e.g., in $T'(D^2W^2)$ implies the derivative by $D^2W^2$, i.e., $T'(D^2W^2) \equiv dT(\xi)/d\xi \big|_{\xi \equiv D^2W^2}$. In order to embed the standard DBI action [1], we need to specify the function $T(\xi)$ to be[5]

$$T(\xi) = \frac{\sqrt{1-c^2\xi}-1}{\xi} = -\frac{c^2}{2\sqrt{\pi}} \sum_{n=1}^\infty \frac{\Gamma\left(n-\tfrac{1}{2}\right)}{n!}(c^2\xi)^{n-1}$$
$$= -\frac{c^2}{2} - \frac{c^4}{8}\xi - \frac{c^6}{16}\xi^2 + \mathcal{O}(\xi^3) \quad . \tag{2.5}$$

Accordingly, the purely bosonic terms of $I_T$ are

$$I_T\big|_{\rm bosons} = \int d^3x \left[ \sqrt{\det(\eta_{ab} + cF_{ab})} - 1 - \frac{c^2}{4}F_{ab}{}^2 \right]$$
$$= \int d^3x \left(\sqrt{1 + \tfrac{c^2}{2}F_{ab}{}^2} - 1 - \tfrac{c^2}{4}F_{ab}{}^2\right) \quad , \tag{2.6}$$

In our new notation here, we use $F_{ab} \equiv (1/2)(\gamma_{ab})^{\alpha\beta}f_{\alpha\beta}$, $f_{\alpha\beta} = (1/2)(\gamma_{ab})_{\alpha\beta}F_{ab}$. The reason of the subtraction of the kinetic term $(c^2/4)F_{ab}{}^2$ in (2.6), or equivalently, the subtraction of $-(c^2/2)W^2$ in (2.3b) is that the self-dual formulation needs no kinetic term of this type [7]. Therefore, the lowest order terms in (2.6) are $F^4$-terms. However, we also stress that our formulation is general enough to accommodate any arbitrary function $T(\xi)$ other than DBI-type action.

---

[5]Of course, the subtraction of $W^2$ from $T(D^2W^2) \cdot D^2W^2$ can be also absorbed into $T(D^2W^2)$ itself. However, the subtraction was made manifest in (2.3b) as a 'reminder'.

For explicitness, we give the $\mathcal{O}(\phi^4)$-terms, namely, the first non-trivial terms quartic in fields, after the kinetic term in $I_T$. They are also of $\mathcal{O}(c^4)$:

$$I_T|_{\phi^4} = \int d^3x \, \frac{c^4}{32} \Big[ -(f_{\alpha\beta}f^{\alpha\beta})^2 + 4if_{\alpha\beta}f^{\alpha\beta}\lambda^\gamma \partial_\gamma{}^\delta \lambda_\delta - 4if^{\alpha\gamma}\lambda_\gamma \partial_\alpha{}^\beta(f_\beta{}^\delta \lambda_\delta)$$
$$+ 4(\lambda^\alpha \partial_\alpha{}^\beta \lambda_\beta)^2 - 2\lambda^2 \partial_{\alpha\beta}\partial^{\alpha\beta}(\lambda^2) \Big] \, . \tag{2.7}$$

There is no more simplification, even though some terms look similar. The basic structure here is the same as in the typical supersymmetric DBI-type actions, such as in 10D [13] or in 3D [12].

The fields $\chi_\alpha$ and $B$ are auxiliary in our system. Suppose temporarily the total lagrangian is $\mathcal{L}_{m^2} + \mathcal{L}_{\mathrm{CS}}$ for simplicity. Then the $\lambda$ and $B$-field equations yield

$$\chi_\alpha \doteq -m^{-1}\lambda_\alpha \, , \quad B \doteq 0 \, . \tag{2.8}$$

Re-substituting these into $\mathcal{L}_{m^2} + \mathcal{L}_{\mathrm{CS}}$, we see this system coincides with the supersymmetric generalization of self-dual system (1.1) as in [7]. Even in our more general system with $\mathcal{L}_T$, the essential structure stays the same, due to the elimination of the kinetic term from $\mathcal{L}_T$, as has been mentioned above.

The $x$-space lagrangian $\mathcal{L}(x)$ can be easily obtained as usual by $\int d^2\theta \, \mathcal{L}(z) = D^2\mathcal{L}(z)|$ applied to $\mathcal{L}_T$, $\mathcal{L}_{m^2}$ and $\mathcal{L}_{\mathrm{CS}}$ in (2.3). Some useful formulae are

$$D^2 W_\alpha = +i\partial_\alpha{}^\beta W_\beta \, , \tag{2.9a}$$

$$D_\alpha D^2(W^2) = +i\partial_\alpha{}^\beta P_\beta \, , \tag{2.9b}$$

$$D_\alpha f_{\beta\gamma} = +\tfrac{i}{2}\partial_{\alpha(\beta}W_{\gamma)} - \tfrac{i}{2}C_{a(\beta}\partial_{\gamma)}{}^\delta W_\delta \, , \tag{2.9c}$$

$$D^2 D^2(W^2) = \tfrac{1}{2}\partial_{\alpha\beta}\partial^{\alpha\beta}(W^2) \, . \tag{2.9d}$$

$$\partial_{\alpha\beta}f^{\alpha\beta} \equiv 0 \, , \tag{2.9e}$$

$$[D^2, D_\alpha] = -2i\partial_\alpha{}^\beta D_\beta \, . \tag{2.9f}$$

As for the invariance of our action under supersymmetry, since all of our actions $I_T$, $I_{m^2}$ and $I_{\mathrm{CS}}$ are given in terms of superfields, the confirmation of the invariance of these actions is manifest. Namely, we can apply the derivative $D_\alpha$ on each $z$-space lagrangians $\mathcal{L}_T(z)$, $\mathcal{L}_{m^2}(z)$ and $\mathcal{L}_{\mathrm{CS}}(z)$, using relationships involving the $D$-operators. The most non-trivial one is that for $\mathcal{L}_T(z)$. To be more specific, when we apply $D_\alpha$ on (2.3b), there arise three sorts of terms: $T''$, $T'$ and $T$-terms depending on the number of derivatives. There will not arise $T'''$-term due to the antisymmetry of indices, and the fermionic dimension $\alpha$, $\beta$, $\cdots$ = 1, 2. For the $T'$-terms, we need a useful lemma

$$D_\alpha \left( \tfrac{1}{2} W^2 \, \partial_{\beta\gamma} \partial^{\beta\gamma} W^2 + i P^\alpha \partial_\alpha{}^\beta P_\beta \right) \Big|$$
$$= -2i \mathcal{L}_0(x) \cdot \partial_\alpha{}^\beta P_\beta |$$
$$+ \partial_{\beta\gamma} \Big[ -\tfrac{1}{2} P_\alpha \partial^{\beta\gamma}(W^2) + \tfrac{1}{2} W^2 \partial^{\beta\gamma} P_\alpha + P^\beta \partial_\alpha{}^\gamma(W^2) - i \delta_\alpha{}^\gamma P^\beta \cdot \mathcal{L}_0(x) \Big] \Big| \quad (2.10)$$

The $T''$-terms can be simplified by the relationship

$$D_\alpha P^\gamma | = i \partial_\alpha{}^\gamma (W^2) | - \delta_\alpha{}^\gamma \mathcal{L}_0(x) \quad , \quad (2.11)$$

By the use of (2.11), the original $T''$-terms can be reduced to $T'$-terms. After certain simplifications among these $T'$-terms, they are combined to produce a total divergence $\partial_\alpha{}^\beta [iT(D^2 W^2) \cdot P_\beta]$ which does not contribute under $\int d^3 x$, as desired.

We can also confirm the $x$-space component lagrangians $\mathcal{L}_T(x)$, $\mathcal{L}_{m^2}(x)$ and $\mathcal{L}_{\text{CS}}(x)$ under the component supersymmetry transformation rule compatible with [8]:

$$\delta_Q V_{\alpha\beta} = -i\epsilon_{(\alpha} \lambda_{\beta)} - \epsilon^\gamma \partial_{\alpha\beta} \chi_\gamma \quad , \quad (2.12\text{a})$$
$$\delta_Q \lambda_\alpha = -\epsilon^\beta f_{\beta\alpha} \quad , \quad (2.12\text{b})$$
$$\delta_Q \chi_\alpha = \epsilon_\alpha B - i\epsilon^\beta V_{\alpha\beta} \quad , \quad (2.12\text{c})$$
$$\delta_Q B = -\epsilon^\alpha \lambda_\alpha + i\epsilon^\alpha \partial_{\alpha\beta} \chi^\beta \quad . \quad (2.12\text{d})$$

Compared with [7], there are certain differences, *e.g.*, (2.12b) does not have a $V$-linear term. Such 'off-shell' differences disappear on-shell under (2.8). However, for more generalized interactions in (2.3), the relationship (2.8) is no longer valid, so that (2.12) is more general than the transformation rule in [7].

It is also convenient to have

$$\delta_Q f_{\alpha\beta} = -\tfrac{i}{2} \epsilon^\gamma \partial_{\gamma(\alpha} \lambda_{\beta)} + \tfrac{i}{2} \epsilon_{(\alpha} \partial_{\beta)}{}^\gamma \lambda_\gamma \quad . \quad (2.13)$$

As usual [8], this rule is obtained by applying the operation $\epsilon^\alpha D_\alpha$, *e.g.*, $\delta_Q V_{\alpha\beta} = -\epsilon^\gamma D_\gamma \Gamma_{\alpha\beta}|$, *etc.* Compared with $I_T$, $I_{\text{CS}}$, the action $I_{m^2}$ is the most non-trivial, because it contains some fields away from the Wess-Zumino gauge. In the course of these invariance confirmations, we need the relationships

$$D_\alpha \Gamma_\beta | = -C_{\alpha\beta} B + i V_{\alpha\beta} \quad , \quad (2.14\text{a})$$
$$D^2 \Gamma_\alpha = +2 W_\alpha + i \partial_{\alpha\beta} \Gamma^\beta \quad , \quad (2.14\text{b})$$
$$f_{\alpha\beta} = -\tfrac{1}{2} \partial_{(\alpha}{}^\gamma V_{\beta)\gamma} \quad . \quad (2.14\text{c})$$

## 3. CS-Term Embedded into 3rd-Rank Field Strength

In the original formulation of self-duality in 3D, the mass parameter $m$ was always supposed to be a constant [7]. However, as in a supereight-brane formulation for massive IIA superstring [14], we can regard the parameter $m$ as an $x$-dependent scalar field $M(x)$, while introducing a multiplier second-rank field $C_{ab}$ constraining $M$ to be constant. Effectively, this is equivalent to introduce the $x$-space lagrangian term proportional to

$$3\epsilon^{abc} C_{ab} \partial_c M = -\epsilon^{abc} M H_{abc} + \text{(total divergence)} \ , \tag{3.1}$$

where $H$ is the field strength of $B$: $H_{abc} \equiv (1/2)\partial_{[a} C_{bc]}$. On the other hand, the original CS-term with the coefficient $m$ is now absorbed into the modified field strength $H$

$$H'_{abc} \equiv \tfrac{1}{2}(\partial_{[a} C_{bc]} + F_{[ab} A_{c]}) = H_{abc} + \tfrac{1}{2} F_{[ab} A_{c]} \ . \tag{3.2}$$

Based on this principle, we can find the following action is equivalent to $I_{m^2} + I_{\text{CS}}$:

$$I' \equiv \int d^3x \left[ +\tfrac{1}{12}\epsilon^{abc} M H'_{abc} - \tfrac{i}{2}(\overline{\lambda}\slashed{\partial}\lambda) + \tfrac{1}{2}M^2 A_a^2 - \tfrac{1}{2}M(\overline{\lambda}\lambda) \right] \ . \tag{3.3}$$

Here we are using the tensor-manifest notation, in which the third-rank field strength looks manifest under the $\epsilon$-tensor. We also use $A_a$ for $V_{\alpha\beta}$, with appropriate scalings of other fields, that put the lagrangian in a more conventional form. We have also eliminated the auxiliary fields $B$ and $\chi$, in the absence of interaction terms. Action $I'$ is invariant under supersymmetry

$$\delta_Q A_a = \ +i(\overline{\epsilon}\gamma_a \lambda) - M^{-1}(\overline{\epsilon}\partial_a \lambda) \ , \tag{3.4a}$$

$$\delta_Q \lambda = \ -i\gamma^a \epsilon M A_a \ , \tag{3.4b}$$

$$\delta_Q C_{ab} = \ +2i(\overline{\epsilon}\gamma_{[a}\lambda) A_{b]} + 2M^{-1}(\overline{\epsilon}\partial_{[a}\lambda) A_{b]} \tag{3.4c}$$

$$= \ -2(\delta_Q A_{[a}) A_{b]} + 4i(\overline{\epsilon}\gamma_{[a}\lambda) A_{b]} \ , \tag{3.4d}$$

$$\delta_Q M = \ 0 \ . \tag{3.4e}$$

A useful relationship for the action invariance under supersymmetry is

$$\delta_Q H'_{abc} = +\tfrac{1}{2}\partial_{[a|} \left[ \delta_Q C_{|bc]} + 2(\delta_Q A_{|b|}) A_{|c]} \right] + (\delta_Q A_{[a|}) F_{|bc]} \ . \tag{3.5}$$

Note that the first term in (3.4d) is the usual routine term cancelling the unwanted bare $A_a$-term in $\delta_Q H_{abc}$ (3.5), while the second term in (3.4d) is the 'net' transformation for $\delta_Q C_{ab}$. Since $M$ is eventually a constant on-shell, there is nothing problematic to have the zero-transformation $\delta_Q M = 0$.

In this formulation with the third-rank field strength, it is more natural to have the CS-term with the scalar field in front, *via* $\epsilon M H$-term. The potential $B_{ab}$ plays a role of lagrange multiplier for constraining $M$ to be $x$-independent quite naturally within the same $\epsilon M H$-term. In this sense, the self-duality in 3D can be more naturally formulated in terms of third-rank field strength $H_{abc}$.

## 4. Supersymmetry Breaking vs. Topology

As a by-product of introducing a tensor multiplet, we consider here a system of supersymmetry breaking and the topological effect of Chern-Simons term. To this end, we temporarily forget about the self-duality, and generalize our previous Abelian vector multiplet to non-Abelian one. In this section, the tensor multiplet is generalized to have kinetic terms.

Consider two multiplets: The Yang-Mills vector multiplet $(A_a{}^I, \lambda^I)$ and the tensor multiplet $(C_{ab}, \chi, \varphi)$. Here the superscript $I$ is for the adjoint representation of a non-Abelian gauge group. The on-shell degrees of freedom are $1+1$ for both multiplets, while the off-shell degrees of freedom are $2+2$ for the vector multiplet as usual, and $C_{ab}(1), \chi(2), \varphi(1)$, so again $2+2$ for the tensor multiplet. Our total action $I_{\text{V,T},f} \equiv \int d^3x \, \mathcal{L}$ consists of the three lagrangians

$$I_{\text{V,T},f} \equiv I_{\text{VK}} + I_{\text{TK}} + I_f \equiv \int d^3x \, (\mathcal{L}_{\text{VK}} + \mathcal{L}_{\text{TK}} + \mathcal{L}_f) \, , \tag{4.1a}$$

$$\mathcal{L}_{\text{VK}} \equiv -\tfrac{1}{4}(F_{ab}{}^I)^2 - \tfrac{i}{2}(\overline{\lambda}^I \gamma^a D_a \lambda^I) \, , \tag{4.1b}$$

$$\mathcal{L}_{\text{TK}} \equiv +\tfrac{1}{12}(\widehat{H}_{abc})^2 - \tfrac{i}{2}(\overline{\chi}\gamma^a \partial_a \chi) + \tfrac{1}{2}(\partial_a \varphi)^2$$
$$= +\tfrac{1}{2}\widehat{H}^2 - \tfrac{i}{2}(\overline{\chi}\gamma^a \partial_a \chi) + \tfrac{1}{2}(\partial_a \varphi)^2 \, , \tag{4.1c}$$

$$\mathcal{L}_f \equiv \tfrac{1}{6}\epsilon^{abc} f(\varphi)\widehat{H}_{abc} - \tfrac{1}{2}f'(\varphi)(\overline{\chi}\chi)$$
$$= +f(\varphi)\widehat{H} - \tfrac{1}{2}f'(\varphi)(\overline{\chi}\chi) \, , \tag{4.1d}$$

The field strength $\widehat{H}_{abc}$ is a modification of $H_{abc} \equiv (1/2)\partial_{[a} C_{bc]}$ defined by

$$\widehat{H}_{abc} \equiv \left( \tfrac{1}{2}\partial_{[a} C_{bc]} + \tfrac{1}{2} F_{[ab}{}^I A_{c]}{}^I - g f^{IJK} A_a{}^I A_b{}^J A_c{}^K \right) - \epsilon_{abc}(\overline{\lambda}^I \lambda^I) \, , \tag{4.2}$$

while $f(\varphi)$ is an arbitrary function of $\varphi$. Thus the usual CS-term in 3D arises naturally in the modified field strength $\widehat{H}$. Compared with the usual modification of such a field strength, we have the $\lambda^2$-term as an extra term. The $\widehat{H}$ is the Hodge dual of $\widehat{H}_{abc}$ defined by

$$\widehat{H} \equiv +\tfrac{1}{6}\epsilon^{abc}\widehat{H}_{abc} \, . \tag{4.3}$$

Note that $\widehat{H}_{abc}$ has the usual CS-term, so that the coefficient $f(\varphi)$ in front of $\epsilon^{abc}\widehat{H}_{abc}$ is to be quantized, when it develops v.e.v., as will be seen shortly. As a special case, a similar lagrangian with $f(\varphi) = \varphi$ with $N = 2$ supersymmetry was given in [15].

Our actions $I_{\rm VK}$, $I_{\rm TK}$, $I_f$, as well as the total action $I_{\rm V,T,}{}_f$ are separately invariant under supersymmetry

$$\delta_Q A_a{}^I = +i(\overline{\epsilon}\gamma_a\lambda^I) \ , \tag{4.4a}$$

$$\delta_Q \lambda^I = -\tfrac{1}{2}\gamma^{ab}\epsilon F_{ab}{}^I \ , \tag{4.4b}$$

$$\delta_Q C_{ab} = -(\overline{\epsilon}\gamma_{ab}\chi) + A_{[a|}{}^I(\delta_Q A_{|b]}{}^I) \ , \tag{4.4c}$$

$$\delta_Q \chi = +\tfrac{1}{6}\epsilon^{abc}\epsilon\,\widehat{H}_{abc} + i\gamma^a\epsilon\,\partial_a\varphi = +\epsilon\widehat{H} + i\gamma^a\epsilon\,\partial_a\varphi \ , \tag{4.4d}$$

$$\delta_Q \varphi = +(\epsilon\chi) \ . \tag{4.4e}$$

As (4.4d) shows, $\widehat{H}$ is similar to the scalar auxiliary field $F$ or $G$ in the chiral multiplet in 4D, which indicates supersymmetry breaking. The presence of the $\lambda^2$-term in the $\widehat{H}$'s simplifies its transformation as

$$\delta_Q \widehat{H}_{abc} = -\tfrac{1}{2}(\overline{\epsilon}\gamma_{[ab}\partial_{c]}\chi) \ , \quad \delta_Q \widehat{H} = +i(\overline{\epsilon}\slashed{\partial}\chi) \ . \tag{4.5}$$

The field equations for $C_{ab}$, $\varphi$ and $A_a$ are respectively

$$\partial_a(\widehat{H} + f(\varphi)) \doteq 0 \ , \tag{4.6a}$$

$$\partial_a^2\varphi + f'(\varphi)\widehat{H} - \tfrac{1}{2}f''(\varphi)(\overline{\chi}\chi) \doteq 0 \ , \tag{4.6b}$$

$$D_b F^{ab\,I} - \epsilon^{abc}(\widehat{H} + f(\varphi))F_{bc}{}^I \doteq 0 \ , \tag{4.6c}$$

where $\doteq$ stands for a field equation.

We now discuss supersymmetry breaking. Consider static solutions: $\varphi = {\rm const.} \equiv \varphi_0$, $\widehat{H} = {\rm const.} \equiv \widehat{H}_0$, and $A_a = 0$, $\chi = 0$. In such a case, (4.6a) is trivially satisfied, while (4.6b) yields the condition

$$f'(\varphi_0)\widehat{H}_0 = 0 \implies \begin{cases} f'(\varphi_0) = 0 \ , & (4.7a) \\ \widehat{H}_0 = 0 \ . & (4.7b) \end{cases}$$

The solution (4.7a) is more interesting, because $\widehat{H}_0$ does not have to vanish, so that supersymmetry can be broken. Accordingly, $\chi$ will be the Nambu-Goldstino, due to $\delta_Q\chi$ in (4.4d). On the contrary, supersymmetry is not broken for (4.7b).

Note that in both of these cases, the value of $\widehat{H}_0$ is 'constant'. Therefore, even though $\mathcal{L}_f$ contains the Chern-Simons term

$$+\tfrac{1}{2}f_0\epsilon^{abc}\left(F_{ab}{}^I A_c{}^I - \tfrac{1}{3}gf^{IJK}A_a{}^I A_b{}^J A_c{}^K\right) \ , \tag{4.8}$$

the coefficient $f_0/2$ is *not* to be quantized. The reason is that any topologically non-trivial configuration for $F_{ab}{}^I \doteq 0$ disturbs the assumption $\widehat{H}_0 = $ const.

The topological configuration arises, when the kinetic lagrangian $\mathcal{L}_{\text{TK}}$ is absent from the system. In such a case, the first term in (4.6a) disappears, and the value of $\widehat{H}$ is not constrained. Eventually, (4.6a) is satisfied, as long as $\varphi = $ const. $\equiv \varphi_0$. Eq. (4.6b) requires $f'(\varphi_0)\widehat{H} = 0$, if $\chi = 0$. Since we need $\widehat{H}$ to be non-trivial, the only appropriate solution is

$$f'(\varphi_0) = 0 \ , \tag{4.9}$$

In other words, the function should be stationary at $\varphi = \varphi_0$. Now eq. (4.6c) allows the non-trivial CS configuration for $F_{ab}{}^I = 0$. In this case, $\mathcal{L}_f$ is simply

$$\mathcal{L}_f \to f_0 \widehat{H} = \tfrac{1}{2} f(\varphi_0) \epsilon^{abc} \partial_a C_{bc} + \tfrac{1}{2} f(\varphi_0) \epsilon^{abc} \left( F_{ab}{}^I A_c{}^I - \tfrac{1}{3} g f^{IJK} A_a{}^I A_b{}^J A_c{}^K \right) \ . \tag{4.10}$$

The first term is a total divergence with no contribution, while the second term is the usual CS-term, whose coefficient $f(\varphi_0)/2$ is to be quantized for a gauge group $G$ with non-trivial homotopy $\pi_3(G) = \mathbb{Z}$:

$$f(\varphi_0) = \tfrac{1}{8\pi} n \quad (n = 0, \ \pm 1, \ \pm 2, \ \cdots) \ . \tag{4.11}$$

Therefore, we have two simultaneous conditions (4.9) and (4.11) for the function $f(\varphi)$. This implies that the minimum or stationary value $f(\varphi_0)$ is to be quantized by (4.11).

To conclude, we have seen that the presence of the kinetic term $\mathcal{L}_{\text{TK}}$ prevents the usual CS-quantization, even though such a CS-term is present in $\mathcal{L}_f$, while the field strength $\widehat{H}$ plays a role of an order parameter for supersymmetry breaking. In the case $\mathcal{L}_{\text{TK}}$ is absent, the minimum or stationary value of $f(\varphi)$ is to be quantized by (4.11).

## 5. Concluding Remarks

In this paper, we have shown in superspace that self-dual $N = 1$ supersymmetric vector multiplet can accommodate a DBI action as interacting terms. Since our polynomial $T(x)$ in (4.1) does not have to be that for DBI action, but an arbitrary polynomial, our result also shows that the self-duality in 3D can have a pretty general form of interaction terms.

The generalized self-duality in 3D needs the field content away from the usual Wess-Zumino gauge. Despite such a complication, we have seen that the supersymmetric invariance of the total action including the 'mass-term' is not spoiled.

There are two important significances of our results in this paper. First, we have shown that self-dual $N = 1$ supersymmetric system in 3D can have non-trivial interaction terms, such as DBI action. Second, the $N=1$ supersymmetric DBI action can be compatible with self-duality in 3D.

As a by-product, we have shown that a third-rank field strength $H'$ leads to a more natural formulation of self-duality in 3D. Following the supereight brane formulation for massive IIA superstring in 10D [14], we have first regarded the mass parameter $m$ as an $x$-dependent scalar field. Next we have introduced the third-rank field strength $H$ dual to the scalar field $M$. In such a formulation, the constraint lagrangian for $\partial_a M = 0$ is naturally embedded into the term $\epsilon^{abc} M H'_{abc}$, while the original CS-term is absorbed into the modified field strength $H'_{abc}$. In other words, this $M \wedge H'$-term plays two roles of both embedding the CS-terms and constraining $\partial_a M = 0$ at the same time. In this sense, this third-rank field strength formulation is much more natural for formulating self-duality in 3D.

Another by-product is the simple mechanism of supersymmetry breaking vs. CS-quantization, both played by the modified third-rank field strength $\widehat{H}$. We have seen that the field strength plays a role of a controlling parameter for supersymmetry breaking for the action $I_{\text{V,T},f}$. When the kinetic terms for the tensor multiplet are absent, we have seen that the CS-term in $\mathcal{L}_f$ requires the usual CS-quantization.

In our formulation, the function $T(D^2 W^2)$ is arbitrary enough for us to combine three different terms. Namely, we have simultaneously the kinetic term $F_{ab}{}^2$, the CS-term $m\epsilon^{abc} A_a F_{bc}$ and the mass term $m A_a{}^2$, extrapolating between the 'topological massive theory' and self-dual theory in 3D. The eminent aspect here is that all of such terms can be treated consistently with manifest supersymmetry all formulated in terms of superfields. Note that the result of DBI action with self-duality is forming just a small subset of infinitely many different theories in 3D presented in this paper.

We are grateful to W. Siegel for important discussions. This work is supported in part by NSF Grant # 0308246.

# References

[1] M. Born and L. Infeld, Proc. Roy. Soc. Lond. **A143** (1934) 410; *ibid.* **A144** (1934) 425; P.A.M. Dirac, Proc. Roy. Soc. Lond. **A268** (1962) 57.

[2] E.S. Fradkin and A. Tseytlin, Phys. Lett. **163B** (1985) 123; A. Tseytlin, Nucl. Phys. **B276** (1986) 391.

[3] R.G. Leigh, Mod. Phys. Lett. **A4** (1986) 2767; C.G. Callan and J.M. Maldacena, Nucl. Phys. **B513** (1998) 198; G.W. Gibbons, Nucl. Phys. **B514** (1998) 603.

[4] S. Gonorazky, C. Nuñez, F.A. Schaposnik, and G. Silva, Nucl. Phys. **B531** (1999) 168; S. Gonorazky, F.A. Schaposnik and G. Silva, Phys. Lett. **449B** (1999) 187; B. Brinne, S.E. Hjelmeland and U. Lindström, Phys. Lett. **459B** (1999) 507.

[5] A.A. Tseytlin, Nucl. Phys. **B501** (1997) 41; J.-H. Park, Phys. Lett. **458B** (1999) 471.

[6] S. Deser and R. Jackiw, Phys. Lett. **139B** (1984) 371.

[7] P.K. Townsend, K. Pilch and P. van Nieuwenhuizen, Phys. Lett. **136B** (1984) 38; Addendum: **137B** (1984) 443.

[8] S.J. Gates, Jr., M.T. Grisaru, M. Roček and W. Siegel, *'Superspace'* (Benjamin/Cummings, Reading, MA 1983).

[9] R. Banerjee, H.J. Rothe and K.D. Rothe, Phys. Rev. **D52** (1995) 3750, `hep-th/9504067`; J.C. Le Guillou and E.F. Moreno, Mod. Phys. Lett. **A12** (1997) 2707, `hep-th/9707210`; M. Gomes, L.C. Malacarne and A.J. da Silva, Phys. Lett. **439B** (1998) 137, `hep-th/9711184`; P.J. Arias, L. Leal and J.C. Perez-Mosquera, Phys. Rev. **D67** (2003) 025020, `hep-th/0206082`; A. Ilha and C. Wotzasek, Phys. Lett. **519B** (2001) 169, `hep-th/0106199`; A. Ilha and C. Wotzasek, Nucl. Phys. **B604** (2001) 426, `hep-th/0104115`; M.A. Anacleto, A. Ilha, J.R.S. Nascimento, R.F. Ribeiro and C. Wotzasek, Phys. Lett. **504B** (2001) 268, `hep-th/0104152`.

[10] E. Harikumar, A. Khare, M. Sivakumar and P.K. Tripathy, Nucl. Phys. **B618** (2001) 570, `hep-th/0104087`; D. Bazeia, A. Ilha, J.R.S. Nascimento, R.F. Ribeiro and C. Wotzasek, Phys. Lett. **510B** (2001) 329, `hep-th/0104098`; P.K. Tripathy and A. Khare, Phys. Lett. **504B** (2001) 152, `hep-th/0009130`; M. Slusarczyk and A. Wereszczynski, Acta Phys. Polon. **B34** (2003) 2623, `hep-th/0204104`.

[11] V.E.R. Lemes, C.L. de Jesus, C.A.G. Sasaki, S.P. Sorella, L.C.Q. Vilar and O.S. Ventura, Phys. Lett. **418B** (1998) 324, `hep-th/9708098`; V.E.R. Lemes, C.L. de Jesus, S.P. Sorella, L.C.Q. Vilar and O.S. Ventura, Phys. Rev. **D58** (1998) 045010, `hep-th/9801021`; M.A.M. Gomes and R.R. Landim, *'Duality and Fields Redefinition in Three Dimensions'*, `hep-th/0405266`.

[12] B. Brinne, S.E. Hjelmeland and U. Lindstrom, Phys. Lett. **459B** (1999) 507, `hep-th/9904175`.

[13] E. Bergshoeff, M. Rakowski and E. Sezgin, Phys. Lett. **185B** (1987) 371.

[14] E. Bergshoeff, M.B. Green, G. Papadopoulos and P.K. Townsend, `hep-th/9511079`; E. Bergshoeff, M. de Roo, M.B. Green, G. Papadopoulos and P.K. Townsend, Nucl. Phys. **B470** (1996) 113, `hep-th/9601150`.

[15] H. Nishino and S.J. Gates, Jr., Int. Jour. Mod. Phys. **A8** (1993) 3371.